# Observation of Dirac Cone Band Dispersion in FeSe Thin Films by Photoemission Spectroscopy

S. Y. Tan[1], Y. Fang[1], D. H. Xie[1], W. Feng[1], C. H. P. Wen[2], Q. Song[2], W. Zhang[1], Q. Y. Chen[1,2], Y. Zhang[1], L. Z. Luo[1], B. P. Xie[2], D. L. Feng[2], X. C. Lai[1*]

[1] Science and Technology on Surface Physics and Chemistry Laboratory, Mianyang 621908, China

[2] Physics Department, Applied Surface Physics State Key Laboratory, and Advanced Materials Laboratory, Fudan University, Shanghai 200433, China

**The electronic structure of FeSe thin films grown on SrTiO₃ substrate is studied by angle-resolved photoemission spectroscopy (ARPES). We reveal the existence of Dirac cone band dispersions in FeSe thin films thicker than 1 Unit Cell below the nematic transition temperature, whose apex are located -10 meV below Fermi energy. The evolution of Dirac cone electronic structure for FeSe thin films as function of temperature, thickness and cobalt doping is systematically studied. The Dirac cones are found to be coexisted with the nematicity in FeSe, disappear when nematicity is suppressed. Our results provide some indication that the spin degrees of freedom may play some kind of role in the nematicity of FeSe.**

    The discovery of high temperature superconductivity in single unit-cell (UC) FeSe film grown on SrTiO$_3$ (STO) substrate has attracted extensively attention [1-9]. A superconducting gap as large as 20 meV was first discovered[1] by scanning tunneling spectroscopy (STS), which was later confirmed[3,4] by angle resolved photoemission spectroscopy (ARPES) measurements. Then, the T$_C$ above 40 K and 100 K in 1-UC FeSe films has been demonstrated by direct transport measurements[8] and in-situ electrical transport measurements[9] respectively. Until now, the superconducting mechanism for single layer FeSe is still unsolved. Another hot debate issue is the driving force of the nematicity for FeSe. Bulk FeSe undergoes a tetragonal to orthorhombic transition at Ts～90 K and at Tc～8 K superconductivity sets in[10]. Clear band splitting around the Brillioun zone corner was first discovered in multi-layer FeSe thin films[4], with a characteristic temperature much higher than the structure transition temperature(T$_s$), which is thought to be caused by short ranged magnetic order. Then similar band reconstruction was also found in bulk FeSe single crystal[11], which was interpreted to be triggered by the electronic nematicity. Although the existence of nematic order in FeSe is by now a well-established experimental fact, its origin remains controversial. Spin, orbital degrees of freedom or their complicated coupling are all possible candidates of the driving force proposed by experimental and theoretical researches[12-22].

    Another intriguing issue in condensed matter physics nowadays is the massless Dirac fermion states in materials, such as graphene[23], topological insulators[24,25], Weyl semimetals[26,27], as well as the parent compound of iron-based superconductors[28-31]. Dirac cone states have been theoretically predicted[28] and experimentally confirmed[29-31] in BaFe$_2$As$_2$ bellow the spin density wave(SDW) transition temperature. It is well established that the formation of Dirac cone states is a consequence of the nodes of the SDW gap by complex zone foldings in bands with different parities. It is even proposed that the coexistence of highly mobile carriers and superconductivity is important for achieving high-T$_c$ superconductors. Surprisingly, ultrafast Dirac cone-like carriers were found in FeSe single crystal by electric transport measurement[32], although no static magnetic order exists in FeSe. Furthermore, a recent DFT calculation[33] predicted a magnetic order named the "pair-checkerboard AFM" as the magnetic ground state of tetragonal FeSe, which can induce Dirac cone band dispersion in FeSe. Thus, it is important to study the possible Dirac cone

states and the related nemativity in FeSe.

In this paper, we report an ARPES investigation of the low-energy electronic states of FeSe thin films grown on STO substrate. We reveal the existence of Dirac cone band dispersions in FeSe thin films thicker than one unit-cell below the nematic transition temperature. The Dirac cones sit at the small spots of high photoemission intensity at the Fermi surface, whose apexes are located -10 meV below Fermi energy, leading to small electron Fermi pockets. The Dirac cone band structure disappears above the nematic transition temperature and barely changes as a function of film thickness. Upon Cobalt doping in multi-layer FeSe films, the nematicity is suppressed significantly and the Dirac cone disappears simultaneously.

High-quality FeSe single crystalline thin films were grown on the $TiO_2$ terminated and Nb-doped $SrTiO_3$ (0.5%wt) substrate with the MBE method following the previous reports[1,4]. After growth, the film was directly transferred from the MBE chamber into the ARPES chamber with typical vacuum of $5 \times 10^{-11}$ mbar. The electron doping is induced by depositing cobalt atoms on the surface of as grown FeSe thin film. ARPES was conducted with 21.2 eV photons from a helium discharge lamp. A SCIENTA R4000 analyzer was used to record ARPES spectra with typical energy and angular resolutions of 10 meV and 0.2°, respectively. A freshly evaporated gold sample in electrical contact with the FeSe sample served to calibrate $E_F$.

The electronic structure of 50 UC FeSe thin film at 30 K is presented in Fig.1. The photoemission intensity map is integrated over a [$E_F$ -10 meV, $E_F$ +10meV] window around the Fermi energy ($E_F$) as shown in Fig.1(a). The observed Fermi surface consists of cross like electron pockets centered at M and oval-shaped hole pockets centered at Γ. The low-energy band structure along the Γ-M direction is shown in Fig.1(b). From the second derivative of the intensity plot with respect to energy in Fig.1(b2), one can clearly observe two hole-like bands contributing to the oval-shaped hole pockets at Γ, where one band cross $E_F$ and the other one with its band top touching $E_F$. The electronic structure around M is much more complicated, mainly consists of two electron-like bands and two hole-like bands, which form four small electron pockets that make the cross shape.

The Fermi surface topology of 50 UC FeSe at 30 K is very similar to those observed in $BaFe_2As_2$ [34] and NaFeAs[35] in their SDW states. Dirac cone band dispersions are discovered at the two very bright spots with high photoemission intensity at the Fermi surface of $BaFe_2As_2$. There are also two bright spots along Γ-M direction at the Fermi surface of 50 UC FeSe film. The low energy electronic structure at 80 K at the brightest spot (labelled as Λ') is present in Fig.1(c1), which clearly shows a Λ-like band dispersion near $E_F$. A distinct Dirac cone like band dispersion can be resolved in Fig.1(c2), when Fig.1(c1) is divided by the Fermi-Dirac function convoluted with the instrument resolution function. The main feature of Λ' point are a hole-like band (labelled as η) and a Dirac cone like band (labelled as χ). The χ band shows clear X-shaped linear dispersion, which shall form a Dirac cone in three dimensional Brillouin zone. The Dirac point ($E_D$) is located -10 meV below $E_F$, forming a small electron pocket which is responsible for the bright spot at the Fermi surface. The momentum distribution curves (MDC) near the Dirac cone is present in Fig.1(c4), the peaks in the MDCs highlight the position of the Λ band. To further illustrate the cone structure, we present the constant-energy maps at different binding energy from $E_F$ to $E_F$-50 meV in Fig.1(d). Four distinct circular Fermi pockets can be resolved centered around M point at $E_F$-50meV binding energy. The sizes of the circular Fermi pockets gradually decrease when binding energy approaching $E_F$, reach their minimum at $E_F$-10 meV and then begin to increase.

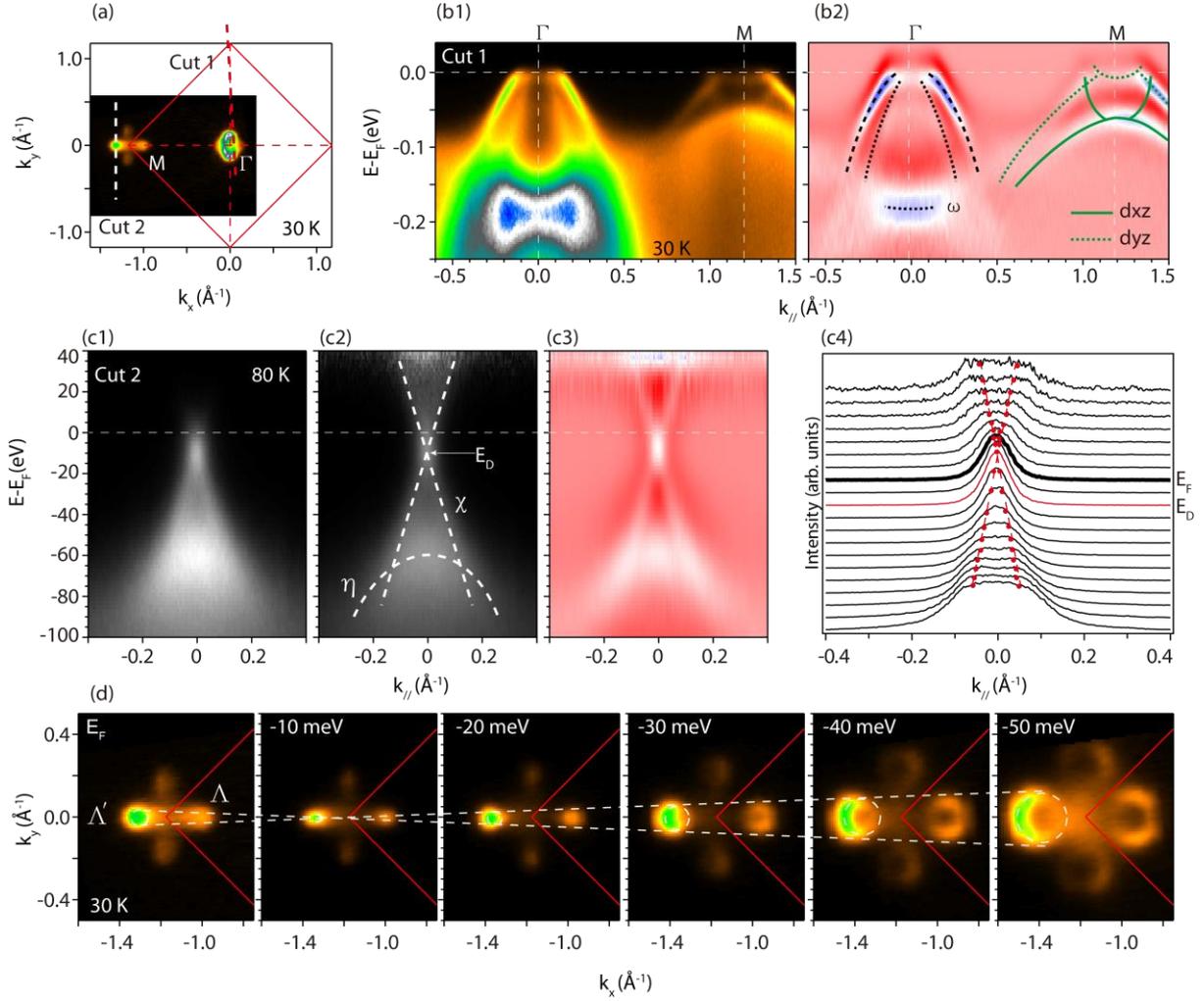

**Fig.1** The Dirac cone band dispersion in 50 UC FeSe thin film. (a) Photoemission intensity map at $E_F$ integrated over [$E_F$ -10 meV, $E_F$ +10 meV] at 30K. (b1,b2), The photoemission intensity and its second derivative of the intensity plot with respect to energy along Γ-M direction(Cut 1) at 30 K. (c1), The photoemission intensity along cut 2 (across the brightest spot in the map) at 80 K. (c2) The ARPES spectra (c1) is divided by the Fermi-Dirac function convoluted with the instrument resolution function, the dashed lines are guides to the eye. (c3) The corresponding second derivative of the intensity plot of c2. (c4) The MDCs around the Dirac cone band, the bold black line refers to $E_F$ and the red line refer to $E_D$. (d) Constant-energy maps at different binding energy from $E_F$ to $E_F$-50 meV.

The Fermi surface topology and band structure around Γ and M points are found to get reconstructed across the nematic transition for both thin film and bulk single crystal FeSe samples. One question immediately follows: Is the Dirac cone band dispersion at Λ point changed across the nematic transition as well? The Fermi surface and band structure below and above the nematic transition temperature ($T_{nem}$=125 K for 50 UC FeSe) are present in Fig.2. The Fermi surface topology at M point changed dramatically from cross-shaped (30 K) to oval-shaped(135 K) across the nematic transition, while Γ point does not change much. We present a series of cuts [from cut 1 to cut 7 in Fig.2(a) around M] at 30 K to illustrate the electronic structure reconstruction more clearly in Fig.2(b). The spectrums taken at the same angles at 135 K are totally different[Fig.2(d)], where cut 4 and cut 7 correspond to Λ and M points.

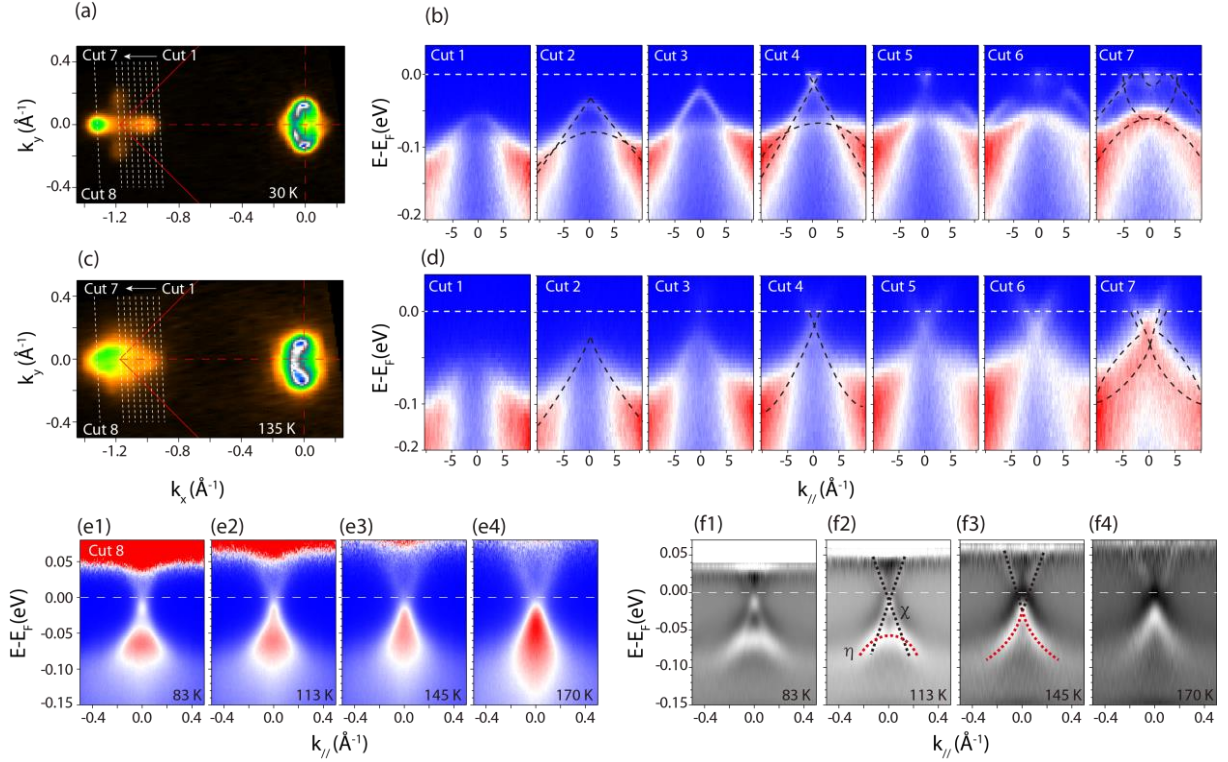

**Fig.2** Temperature dependence of the band structure for 50 UC FeSe thin film. (a) Photoemission intensity map at $E_F$ integrated over [$E_F$-10 meV, $E_F$+10 meV] at 30K. (b) The band structure evolution near M region from cut 1 to cut 7 in (a). (c) Photoemission intensity map at $E_F$ integrated over [$E_F$-10 meV, $E_F$+10 meV] at 135 K. (d) The band structure evolution near M region from cut 1 to cut 7 in (c). (e1-e4) Temperature dependence of the band structure at Λ point after division by the Fermi-Dirac function. (f1-f4) The corresponding second derivative of the intensity plot of (e1-e4).

The band structure evolution at Λ' point(cut 8) as a function of temperature is shown in Fig.2(e) after division by the Fermi-Dirac function. Distinct band structure reconstruction can be resolved in the corresponding second derivative of the intensity plot in Fig.2(f). Below the nematic transition temperature at 113 K and 83 K, the main feature at Λ' point is a hole-like band η and a Dirac cone like band χ as illustrated in Fig.2(f2). At 145 K and 170 K above the nematic transition temperature, it seems to be an electron-like band and a hole-like band intersect around $E_F$. The X like band dispersion which exist at the nematic state disappears above the nematic transition temperature, changes to an conventional electron like band which cross $E_F$ to form a bit larger electron Fermi pocket. Recently, an electric transport measurement under magnetic fields reported possible Dirac cone-like ultrafast electrons in FeSe single crystals[32]. By employing ab initio mobility spectrum analysis, a remarkable reduction in carrier number and an enhancement in carrier mobility were simultaneously observed below 120 K higher than the structural transition temperature($T_s$=90 K) of FeSe single crystal. The previous transport result can be well explained by our ARPES data. Above the nematic transition temperature($T_{nem}$=125 K for 50 UC FeSe), a conventional electron like band with a larger Fermi surface exist at the Λ' point. Below the nematic transition temperature, the electron like band changes to form a Dirac cone like band with a smaller Fermi surface (reduction in carrier number) and ultrafast carrier mobility(enhancement in carrier mobility). Based on the transport and ARPES experimental results, one may conclude that the Dirac cone band dispersion has close relationship with the nematic state in FeSe.

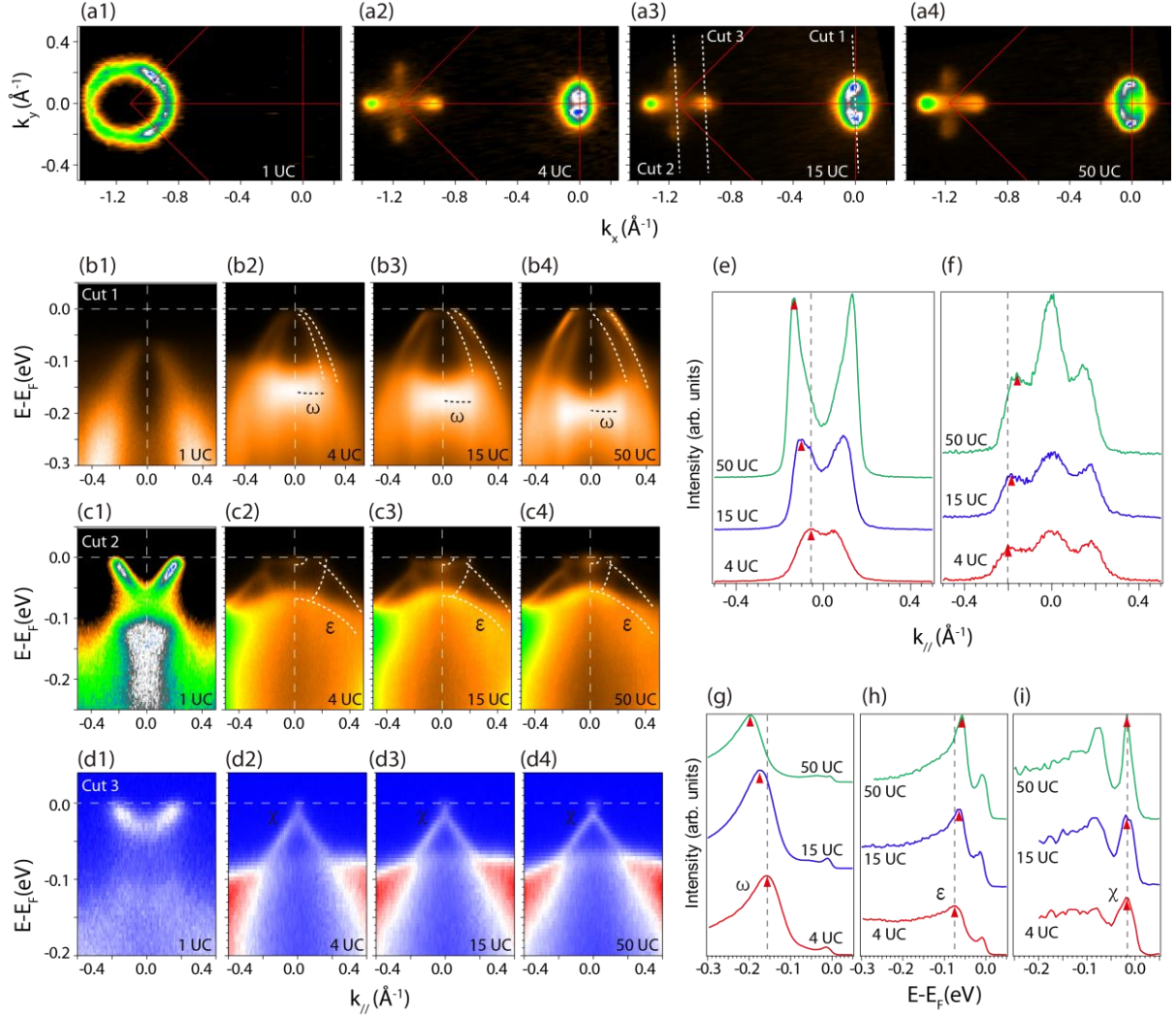

**Fig.3** Thickness dependence of the band structure for FeSe thin films at 30 K. (a) Thickness dependence of photoemission intensity map at $E_F$ integrated over [$E_F$ -10 meV, $E_F$ +10 meV]. (b,c,d) Thickness dependence of the band structure around Γ(b1-b4) (cut1 in a3), M (c1-c4) (cut2 in a3) and Λ point (d1-d4) (cut3 in a3) respectively. (e, f) Thickness dependence of the MDCs at $E_F$ around Γ and M points respectively. (g, h, i) Thickness dependence of the EDCs at k=0 around Γ, M and Λ points respectively.

The nematic transition temperature has been found to decrease with increasing thickness in multi-layer FeSe films[4], which are 165 K, 145 K and 125 K respectively for 4 UC, 15 UC and 50 UC FeSe films. We present thickness dependence of the electronic structure for FeSe thin films at 30 K in Fig.3 to track the evolution of the Dirac cone band dispersion. The superconducting 1 UC FeSe film ($T_c$～50 K) consists of only electron Fermi pocket at M point[ Fig.3(a1)], which shows no Dirac cone like band at Λ point[ Fig.3(d1)]. The Fermi surface topologies of FeSe films thicker than 2 UC are very much alike[ Figs.3(a2)-(a4)], while the low energy bands at Γ and M points change slightly with increased thickness. The spectral weight at Γ point is contributed by two hole-like bands for multi-layer FeSe films[Figs.3(b2)-(b4)]. Their band tops touch $E_F$ in 4 UC film and move upwards with increased thickness, and the hole pockets area increase slightly. The Momentum distribution curves around $E_F$ at Γ point are shown in Fig.3(e), in which the peaks represent the Fermi crossing($k_F$) of the hole band. One can find that the Fermi crossing becomes larger with increased thickness. Moreover, a nearly flat band named ω moves downwards with increased thickness, which is highlighted by the peaks of the EDCs at Γ point in Fig.3(g).

The band structure near M is more complicated, with several bands crossing to give four small electron pockets that make the cross shape. The Fermi crossing of the electron bands become smaller with increased thickness, which are highlighted by the peaks of the MDCs around $E_F$ at M point in Fig.3(f). Moreover, the top of a hole-like band labeled ε moves upwards with increased thickness, which is highlighted by the peaks of the EDCs at M point in Fig.3(h). In contrast to the distinct band evolutions at Γ and M points, the Dirac cone band dispersion at Λ point barely changes as shown in Fig.3(d). The EDCs at $k_F$=0 around Λ point are shown in Fig.3(i), in which the peaks near $E_F$ represent the position of the Dirac point. The Dirac point shows no obvious change within our system resolution.

Theoretically, it has been predicted that nonmagnetic impurities do not affect the Dirac-cone states. Previous studies on the Nernst effect in $Eu(Fe_{1-x}Co_xAs)_2$ [36] and magnetotransport in $Ba(Fe_{1-x}TM_xAs)_2$ (TM = Co, Ni, Cu)[37] indicate that the influence of the Dirac fermion on electronic transport is greatly suppressed by substitution with other 3d transition metals. We used cobalt as dopant and measured the doping dependence of the electronic structure for 4 UC FeSe films, which is shown in Fig.4.

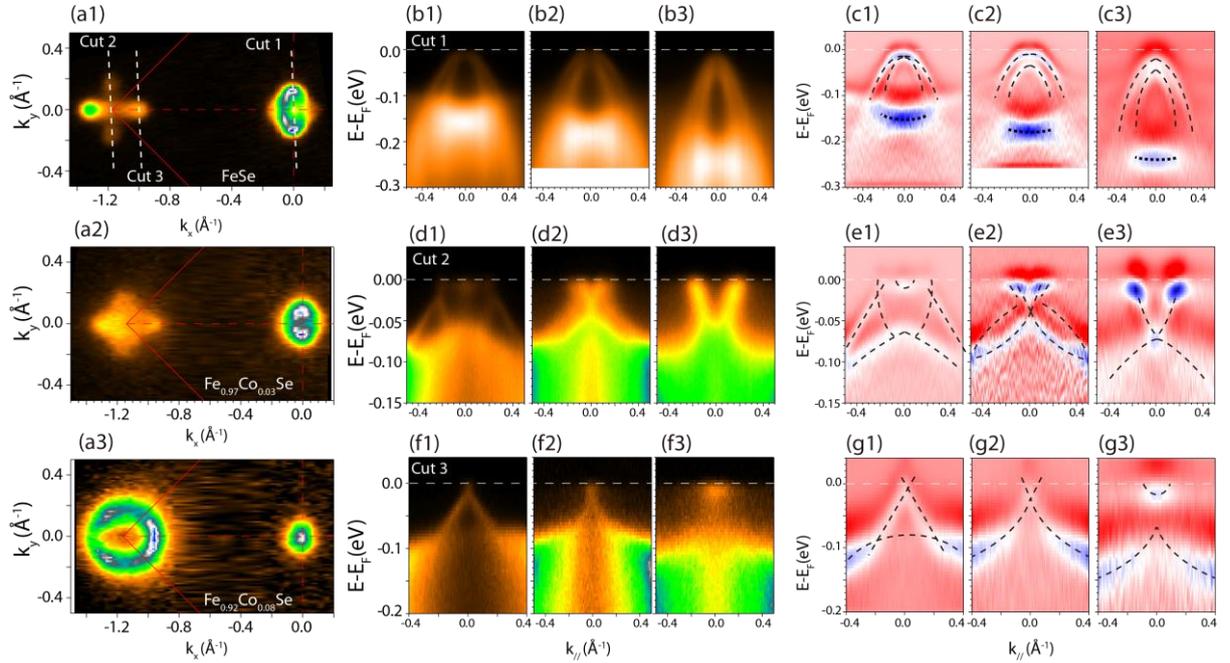

**Fig.4** Doping dependence of the band structure for 4 UC FeSe thin films. (a) Doping dependence of the photoemission intensity map at $E_F$ integrated over [$E_F$-10 meV, $E_F$+10 meV]. (b) Doping dependence of the photoemission intensity and its corresponding second derivative of the intensity plot around Γ point(cut# 1 in a1).(c) Doping dependence of the photoemission intensity and its corresponding second derivative of the intensity plot around M point(cut# 2 in a1). (d) Doping dependence of the photoemission intensity and its corresponding second derivative of the intensity plot around Λ point (cut# 3 in a1).

The Fermi surface topology of 4 UC FeSe changes dramatically upon cobalt doping. With about 3% cobalt doping, the cross shaped electron pockets around M in undoped films[Fig.4(a1)] change to two intersecting elliptical shaped pockets[Fig.4(a2)], which agrees well with the theoretical predicated Fermi surface of ion based superconductor in the paramagnetic state[38]. When the cobalt doping level increased to 8%, a nearly circular electron Fermi pocket emerges at M point[Fig.4(a3)] which is similar to the Fermi surface of 1 UC FeSe[Fig.3(a1)]. As shown in Figs.4(b) and 4(c), the spectral weight at the zone centre near $E_F$ is contributed by two hole-like bands. The two band tops touch $E_F$ in the undoped 4UC film; with increasing cobalt doping, the hole-like bands move downwards, and the hole pockets area decrease.

The band topology changes significantly at M point, which is shown in Figs.4(d) and 4(e). The band structure of 3% cobalt doped FeSe film at M point [Fig.4(e2)] is very similar to the structure at high temperature [Cut 7 in Fig.2(d)] above the nematic transition, in which the nematic state is suppressed by cobalt doping and temperature in the former and later case, respectively. With 8% cobalt doping, one electron-like band which forms the circular Fermi pocket and a hole-like band located beneath the electron band can be resolved. The band structure of 8% cobalt doped FeSe is very similar to 1UC FeSe, but still have distinct differences. The bottom of the electron band intersect the top of the hole band in the 8% cobalt doped film at M point[Fig.4(e3)], while there is a gap between the two bands for 1 UC FeSe[Fig.3(c)]. At Γ point , there is only one hole like band with its top 80 meV below $E_F$ in 1 UC FeSe[Fig.3(b1)], while two hole like bands with their tops close to $E_F$ can be resolved in 8% cobalt doped film[Fig.4(c3)].

The band structure evolution at Λ point is shown in Figs.4(f) and 4(g). With about 3% cobalt doping, the Dirac cone band dispersion in Fig.4(g1) changes to a conventional parabolic electron band in Fig.4(g2). Again, the band structure of 3% cobalt doped FeSe at Λ point is similar to the one at high temperature in Fig.2(f3). Similar to iron-arsenide parent compounds, the nematic state in FeSe film can be suppressed by substitution with other 3d transition metals(which cause electron or hole doping). Consequently, the Dirac cone states disappear when the nematic state is suppressed.

In iron-arsenide compounds, the parent state is a collinear antiferromagnet. A theoretical research[28] found that the combination of physical symmetry and the topology of the band structure naturally stabilizes a gapless SDW ground state with Dirac nodes in undoped FeAs compound. Then an angle-resolved photoemission spectroscopy study[30] revealed the existence of Dirac cones in the electronic structure of $BaFe_2As_2$ below the spin-density-wave temperature, which are responsible for small spots of high photoemission intensity at the Fermi level. Moreover, both electron and hole Dirac cone states are confirmed by magnetoresistance in $BaFe_2As_2$, where transverse magnetoresistance develops linearly against the magnetic field at low temperatures[31].

Similar to $BaFe_2As_2$, a recent DFT calculation[33] predicted a magnetic order named "pair-checkerboard AFM" as the magnetic ground state of tetragonal FeSe, which can induce Dirac cone band dispersion in FeSe. Besides, ultrafast Dirac cone-like carriers were found in FeSe single crystal by electric transport measurement[32], although no static magnetic order exists in FeSe. Our ARPES data reveal that the Fermi surface of FeSe films in the nematic state is similar to $BaFe_2As_2$ in the SDW state, and similar Dirac cone band dispersions are found to be located at two small spots of high intensity. The Dirac cone band dispersion disappears when the nematic state is suppressed by increased temperature or cobalt electron doping. All indications show that there may have some close relation between the nematic state of FeSe and the SDW state of iron-arsenide parent compounds, although no static magnetic order exists in FeSe.

In summary, We reveal the existence of Dirac cone band dispersions in FeSe thin films thicker than 1 Unit Cell below the nematic transition temperature. The Dirac cones sit at the small spots of high photoemission intensity at the Fermi surface, whose apex are located -10 meV below Fermi energy, leading to small electron Fermi pockets. The Dirac cone band structure disappears above the nematic transition temperature and barely changes as a function of film thickness. Upon cobalt doping in multi-layer FeSe films, the nematicity is suppressed significantly and the Dirac cone disappears simultaneously. The Dirac cone states show close relationship with the nematicity, which are predicted to exist in the antiferromagnetic ground state of FeSe similar to the Dirac cones found in $BaFe_2As_2$ in the SDW state. Our results provide some indication that the magnetic degrees of freedom may play some kind of role in the nematicity of FeSe, at least strong interplay between the magnetic and orbital degrees may exist in FeSe.

We gratefully acknowledge helpful discussions with Prof. Z. Y. Lu, Dr. L. Huang and Dr H. C. Xu. This work is supported by the Foundation of President of China Academy of Engineering Physics (Grants No. 201501037) and by the National Science Foundation of China (Grants No. 11504341).

---